\documentclass[letterpaper, 10 pt, conference]{ieeeconf}  % Comment this line out
\IEEEoverridecommandlockouts                              % This command is only
\usepackage{amsmath,amsthm,amssymb,mathtools,bbm,dsfont,physics}

\usepackage[hidelinks]{hyperref}
\usepackage[capitalize]{cleveref}

\usepackage{xcolor}

\usepackage{comment}

\usepackage{etoolbox}
\newbool{booltrue} 
\booltrue{booltrue}
\newbool{boolfalse} 
\boolfalse{boolfalse}

\newtheorem{theorem}{Theorem}
\newtheorem{lemma}[theorem]{Lemma}

\newtheorem{proposition}[theorem]{Proposition}

\newtheorem{assumption}{Assumption}
\newtheorem*{remark}{Remark}

\renewcommand{\d}{\mathrm{d}}
\newcommand{\reals}{\mathbb{R}}
\newcommand{\naturals}{\mathbb{N}}
\DeclareMathOperator{\Id}{Id}
\DeclareMathOperator{\argmin}{argmin}

\newcommand{\diracDelta}[1]{\delta_{#1}}
\newcommand{\pushforward}[1]{#1_{\#}}

\newcommand{\setPlans}[2]{\Gamma(#1,#2)}
\newcommand{\setOptimalPlans}[2]{\Gamma_o(#1,#2)}
\newcommand{\wassersteinDistance}[2]{W(#1,#2)}

\newcommand{\Pp}[2]{\mathcal{P}_{#1}(#2)}

\newbool{finalsubmission} 
\boolfalse{finalsubmission} % Uncomment to include all proofs (arxiv)
\ifbool{finalsubmission}{\def\widthfigures{7.8cm}}{\def\widthfigures{8.2cm}}
\ifbool{finalsubmission}{\AtEndEnvironment{figure}{\vspace{-0.1cm}}}{\AtEndEnvironment{figure}{\vspace{-0.1cm}}}
\ifbool{finalsubmission}{
\setlength{\abovecaptionskip}{0pt}
\setlength{\belowcaptionskip}{0pt}
}

\title{\LARGE \bf
Modeling of Political Systems using Wasserstein Gradient Flows
}

\author{Nicolas Lanzetti, Joudi Hajar, and Florian Dörfler% <-this % stops a space
\thanks{The first two authors contributed equally to this work.}
\thanks{This work was supported by the Swiss National Science Foundation under NCCR Automation, grant agreement 51NF40\_180545.}% <-this % stops a space
\thanks{N. Lanzetti, J. Hajar, F. Dörfler are with the Automatic Control Laboratory, Department of Electrical Engineering and Information Technology, ETH Zürich, 8003 Zürich,  Switzerland, {\tt\small \{lnicolas,jhajar,doerfler\}@ethz.ch}}%
}

\begin{document}

\maketitle
\thispagestyle{empty}
\pagestyle{empty}

%%%%%%%%%%%%%%%%%%%%%%%%%%%%%%%%%%%%%%%%%%%%%%%%%%%%%%%%%%%%%%%%%%%%%%%%%%%%%%%%
\begin{abstract}
The study of complex political phenomena such as parties' polarization calls for mathematical models of political systems. In this paper, we aim at modeling the time evolution of a political system whereby various parties selfishly interact to maximize their political success (e.g., number of votes). More specifically, we identify the ideology of a party as a probability distribution over a one-dimensional real-valued ideology space, and we formulate a gradient flow in the probability space (also called a Wasserstein gradient flow) to study its temporal evolution.
We characterize the equilibria of the arising dynamic system, and establish local convergence under mild assumptions. 
We calibrate and validate our model with real-world time-series data of the time evolution of the ideologies of the Republican and Democratic parties in the US Congress. Our framework allows to rigorously reason about various political effects such as parties' polarization and homogeneity. Among others, our mechanistic model can explain why political parties become more polarized and less inclusive with time (their distributions get ``tighter''), until all candidates in a party converge asymptotically to the same ideological position.
\end{abstract}

%%%%%%%%%%%%%%%%%%%%%%%%%%%%%%%%%%%%%%%%%%%%%%%%%%%%%%%%%%%%%%%%%%%%%%%%%%%%%%%%
\section{Introduction}
%Mathematical political science is gaining more attention among researchers. 
%Recent years have witnessed considerable efforts to understand the dynamics of political competition as well as the positions politicians and people are opting for~\cite{Estimating, conradie2021modelling, Electoral,new2022}.
In American politics, it is puzzling that while most of the American people have moderate opinions about main issues tackled by politicians~\cite{fiorina2008political,hetherington2009,hill2015}, the Republican and Democratic parties are taking positions that are far from the public's moderate ideology and that are becoming increasingly polarized~\cite{mccarty2016,abramowitz2005,baldassarri2008}.
For instance, \cite{fiorina2008political} shows that (i) middle of the road positions are predominant in the public's ideology and that (ii) there is little to no increase in mass polarization.
However, at the same time, the ideological overlap between parties is decreasing, and parties are getting more and more polarized.
In a society designed for political representation, why are politicians taking more extreme positions when the majority of the public opts for centrist positions?

This apparent contradiction calls for mechanistic mathematical models to help politicians and voters better understand their socio-political positions, and possibly optimize for their acts and decisions.
The goal of this paper is to develop a mathematical model for the ideologies of political parties. Specifically, we leverage the theory of gradient flows in probability spaces to formulate a dynamic model for the ideology evolution of political parties. In contrast with existing literature, where a party is usually lumped in its \emph{average} ideological position, our approach allows us to consider the parties' \emph{full ideological distributions}. This way, we can study various political effects such as polarization, homogeneity, and inclusiveness.

\ifbool{finalsubmission}{\paragraph{Related Literature}}{\subsection{Related Literature}}
\label{sec:intro:related literature}

%Our work lies at the interface of mathematical models for political systems and gradient flows in the Wasserstein space.
Political systems are usually modeled with a utility maximizing approach~\cite{johnson1989formal,gersbach2017,krasa2016political}, which finds its root in the Downsian model.
In the Downsian model, competition among utility-maximizing parties is modeled in a one-dimensional space, representing the ideological position of each party (i.e., negative values are left positions and positive values are right positions).
Each candidate of a political party rationally opts for the policy that maximizes their utility, and, given the policy announcement of candidates, voters maximize their expected utility.
The Downsian model predicts that the positions of two competing parties reach consensus, converging to the median of voters' positions.

The Downsian model does not capture empirically observed phenomena such as polarization, and as such it has undergone many extensions and revisions. 
For instance, if the ideology space is multi-dimensional, then parties can focus on ``orthogonal'' political issues, which prevents them from converging to the same ideological position. 
In \cite{gersbach2017} instead, parties and citizens maximize their quadratic preferences and follow Markov strategies. The model predicts convergence to an alternance of policies.
Finally, the utility-maximizing dynamic model in \cite{krasa2016political} explains why and how parties adopt non-moderate policies.

More recently, \cite{Yang2020} proposed a satisficing dynamical model to study polarization of political parties. This model, based on studies showing that people tend to be non-maximizers~\cite{Maximizing}, does not assume that voters maximize their utility, but rather that they opt for candidates who are ``good enough''. Then, parties opportunistically adjust the average of their ideology to maximize their number of votes. Among others, this model explains political polarization well. 

Our work considers modeling of political systems, specifically parties' positioning over an ideology. We probabilistically formalize the parties' utility maximization problem as a gradient flow in the probability space, also called Wasserstein gradient flow.
Wasserstein gradient flows were pioneered by~\cite{jordan1998}, who showed that the Fokker-Plank equation can be considered as a gradient flow in the space of probability distributions endowed with the Wasserstein distance, a distance between probability distributions based on the theory of optimal transport~\cite{Villani2007,Santambrogio2015}. 
The intuition was then extended and formalized in~\cite{Ambrosio2008}, with a whole theory of gradient flows in metric spaces and its specialization to the probability space. For an introduction to Wasserstein gradient flows, we refer to~\cite{santambrogio2017}.
Initially, Wasserstein gradient flows mainly found application in the theory of partial differential equations: Many partial differential equations can be seen, and therefore studied, as Wasserstein gradient flows; e.g., see~\cite{otto1996,otto2001}.
More recently, Wasserstein gradient flows also found application in machine learning~ \cite{mei2019,bunne2021,chewi2020,chizat2018}, reinforcement learning~\cite{zhang2018,richemond2017}, and, more generally, optimization theory~\cite{arbel2019,salim2020,lanzetti2022firstorder}.

\ifbool{finalsubmission}{\paragraph{Contributions}}{\subsection{Contributions}}\label{sec:intro:contributions}
Motivated by the satisficing dynamical model of~\cite{Yang2020}, we provide a model of political systems which accounts for the parties' full ideological distribution, and not only for the average position of its candidates.
More specifically, our contributions is threefold. 
First, we formulate a Wasserstein gradient flow to model the dynamics of the ideological distributions of political parties aiming at maximizing their political success. 
Second, we study the arising equilibria and their convergence properties. 
Finally, we validate our model with data from the US Congress~\cite{dataa}.

\ifbool{finalsubmission}{\paragraph{Organization}}{\subsection{Organization}}\label{sec:intro:organization}
The remainder of this paper is organized as follows. 
In~\cref{sec:background}, we review the model of~\cite{Yang2020}. In~\cref{sec:ourmodel}, we extend it to account for the parties' full ideological distributions, and study its theoretic properties. In~\cref{sec:results}, we perform numerical simulations and validate our model. Finally, \cref{sec:conclusion} draws the conclusions of this paper. 
All proofs are relegated to the appendix.
\section{A Satisficing Dynamical Model}\label{sec:background}

In this section, we review the satisficing dynamical model from~\cite{Yang2020}. We present the model in~\cref{subsec:background:model}, and study its equilibria in~\cref{subsec:background:theory}.

%In~\cref{subsect:background}, we review the satisficing dynamical model from~\cite{Yang2020}. In~\cref{subsect:ourmodel}, we extend it to capture the full ideological distributions of political parties. Finally, in~\cref{subsect:theory}, we study the convergence properties of the arising equilibria. 

\subsection{The Satisficing Dynamical Model from~\cite{Yang2020}}\label{subsec:background:model}
Empirical research has demonstrated that the US political spectrum is well captured by a one-dimensional real-valued \emph{ideology space}: Left positions (i.e., negative values on the real line) represent liberals and right positions (i.e., positive values on the real line) represent conservatives~\cite{Yang2020}. 
Each party is modeled by the \emph{average} ideological position of its candidates in this one-dimensional space, denoted here by $y_i\in\reals$.
Surveys show that the public's ideology has a unimodular distribution, with a peak at centrist positions, which can be well approximated by a Gaussian distribution $\rho(x)$~\cite{Yang2020}. Without loss of generality, we assume that $\rho$ is zero-mean and has the standard deviation $\sigma_0\in\reals_{>0}$.

In constrast to the Downsian model, voters do \emph{not} maximize their utility, but rather opt for the party with which they are satisfied. Should they be satisfied with more than one party, they vote randomly for one of them. Satisfaction is measured via a so-called satisficing function $s_i(d_i)$, where $d_i=\abs{x-y_i}$ is the ideological distance between the voter and party $i$. The semantics is as follows: $s_i(d_i)$ is the probability that a voter with ideological position $x$ is satisfied with party $i$ (which has the average ideological position $y_i$).
When the distance between the voter's position $x$ and the party's position $y_i$ increases, the probability of being satisfied with the party decreases according to 
\begin{equation}\label{eq:background:satisficing}
    s_i(d_i)\coloneqq \exp\left(-\frac{d_i^2}{2\sigma_i^2}\right),
\end{equation}
where $\sigma_i\in\reals_{>0}$ represents the tolerance that voters have to parties with different ideologies than theirs. Henceforth, we will assume that all $\sigma_i$ are identical; i.e.,  $\sigma_i=\sigma\in\reals_{>0}$.
If $\sigma$ is large, more voters with positions far from party $i$'s average ideological position are likely satisfied with the party and might vote for it.
Accordingly, in the simplified case of two parties, a voter opts for party $1$ if
\begin{itemize}
    \item it is satisfied with party 1 only, happening with probability $s_1(d_1)(1-s_2(d_2))$; or
    \item it is satisfied with both parties, happening with probability $s_1(d_1)s_2(d_2)$, and randomly decides to vote for party 1, happening therefore with probability $\frac{1}{2} s_1(d_1)s_2(d_2)$.
\end{itemize}
Thus, the probability that a voter at position $x$ votes for party 1 is
\begin{equation}\label{eq:background:probability}
    p_1(x|y_1,y_2)=s_1(d_1)(1-s_2(d_2))+\frac{1}{2} s_1(d_1)s_2(d_2),
\end{equation}
and the expected total number of voters for party $1$ is
\begin{equation}\label{eq:background:Vtilde}
    \tilde{V_1}(y_1,y_2)
    \coloneqq 
    \mathbb{E}_{\rho}[p_1(x|y_1,y_2)]
    =
    \int_{\mathbb{R}} p_1(x|y_1,y_2)\d\rho(x),
\end{equation} 
where $\d\rho(x)=\rho(x)\d x$.
Of course, all expressions for party 2 are symmetric.

The model in~\cite{Yang2020} assumes that each party opportunistically aims at maximizing its number of votes $\tilde V_1$. Thus, the continuous-time evolution of each party's average ideological position is captured by a gradient flow. Namely, each party moves in the direction that increases its number of votes, at a speed proportional to the potential gain:
\begin{equation}\label{eq:background:dynamic system}
\begin{aligned}
     \dot y_1(t) &= k \nabla_{y_1}\tilde{V_1}(y_1(t),y_2(t))
     &&
     y_1(0)=y_{1,0},
     \\
     \dot y_2(t) &= k \nabla_{y_2}\tilde{V_2}(y_1(t),y_2(t))
     &&
     y_2(0)=y_{2,0},
\end{aligned}
\end{equation} 
where $y_{1,0}, y_{2,0}\in\reals$ are given initial conditions and $k\in\reals_{>0}$ is a positive constant (determined from empirical data).

\subsection{Theoretic Analysis}\label{subsec:background:theory}
Interestingly, in some cases, \eqref{eq:background:dynamic system} predicts that parties do \emph{not} converge to the same ideological position, but rather polarize and converge to asymmetric positions.
To identify in which configurations polarization is an equilibrium, we formulate the following assumption on the parameters of the system: 

\begin{assumption}[Adapted from~\cite{Yang2020}]\label{assumption:parameters}
We have $\sigma/\sigma_0<\sigma_c$, where $\sigma_c\approx 0.807$ is the unique real-valued root of $3\sigma_c^6+5\sigma_c^4-3\sigma_c^2-1=0$.
\end{assumption}

\cref{assumption:parameters} is satisfied whenever voters are not \emph{too tolerant}; i.e., $\sigma$ is sufficiently small compared to $\sigma_0$. Then, parties' polarization is an equilibrium if and only if~\cref{assumption:parameters} holds true. Otherwise, a consensus is reached. 

\begin{proposition}[Adapted from~\cite{Yang2020}]\label{prop:background}
Let~\cref{assumption:parameters} hold.
Then, the dynamic system~\eqref{eq:background:dynamic system} admits three equilibria:
\begin{itemize}
    \item the unstable symmetric equilibrium $y_1^\ast=y_2^\ast=0$; 
    \item two locally asymptotically stable asymmetric equilibria $y_1^\ast=-y_2^\ast=y^\ast$ with 
    \begin{equation}\label{eq:background:asymmetric equilbrium}
        y^\ast
        =\pm\sigma\sqrt{\frac{\sigma^2+\sigma_0^2}{\sigma^2+2\sigma_0^2}\ln\left(\frac{(\sigma^2+\sigma_0^2)^3}{4\sigma^4(\sigma^2+2\sigma_0^2)}\right)}.
    \end{equation}
\end{itemize}
Moreover, if~\cref{assumption:parameters} does not hold, then $y_1^\ast=y_2^\ast=0$ is the only equilibrium, and it is asymptotically stable.
\end{proposition}

\begin{figure}[b]
    \centering
    \includegraphics[width=\widthfigures]{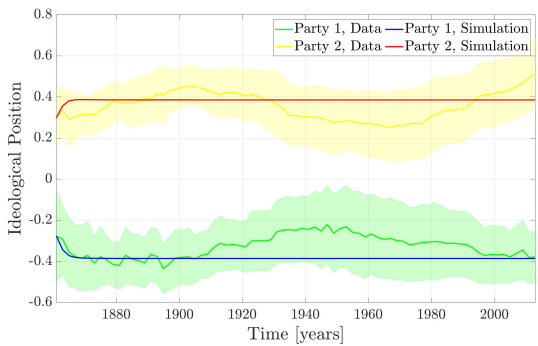}
    \caption{Comparison of the model presented in~\cite{Yang2020} with data. In green and yellow: average plus/minus standard deviation of the ideological distribution of each party, using the data from~\cite{dataa}. In red and blue: predictions of the model using the fitted parameters (i.e., $k=2.54$, and $\sigma=0.4$).}
    \label{fig:original}
\end{figure}

The proof is deferred to the appendix.
In words,~\cref{prop:background} asserts that if voters are not too tolerant (i.e.,~\cref{assumption:parameters} holds true), then parties' polarization is a locally asymptotically stable equilibrium, while the outcome of the Downsian model (i.e., both parties sharing the public's ideology, namely $y_1^\ast=y_2^\ast=0$) is an unstable equilibrium. 
If the tolerance $\sigma$ is increased so that~\cref{assumption:parameters} is violated, then the system undergoes a pitchforck bifurcation, and $y_1^\ast=y_2^\ast=0$ is the unique (asymptotically stable) equilibrium. In this case, both parties asymptotically converge to the average of the public's ideology, as predicted by the Downsian model.
The model was validated with data from the US Congress~\cite{dataa}; see~\cref{fig:original}. 

\section{Model}\label{sec:ourmodel}

We now present our model. In~\cref{subsec:ourmodel:model}, we extend the model from~\cite{Yang2020}, reviewed in~\cref{sec:background}, to capture the full ideological distributions of political parties. In~\cref{subsec:ourmodel:theory}, we study the convergence properties of the arising equilibria. We conclude with some discussion in~\cref{subsec:ourmodel:discussion}.

\subsection{A Distributional Model}\label{subsec:ourmodel:model}
The model in~\cite{Yang2020} represents the position of party $i$ as a single point $y_i\in\mathbb{R}$ on the ideology space. Since parties are usually heterogeneous (indeed, not all candidates share the same ideological position), we extend the model to account for the \emph{full ideological distribution}, and therefore represent the position of party $i$ as a \emph{probability distribution} over the real line. We denote it by $\mu_i\in\Pp{2}{\mathbb{R}}$, where $\Pp{2}{\mathbb{R}}$ is the space of probability distributions over the real line with finite second moment.
For instance, a ``tight'' distribution (e.g., Gaussian with low variance) suggests that a party is quite homogeneous around its average ideological position, and culminates in a \emph{delta distribution} $\delta_{\bar y}$, indicating that the party is homogeneous and all candidates share the ideological position $\bar y\in\reals$.
Conversely, a ``diffused'' distribution (e.g., uniform with high variance) models a heterogeneous party, with very different ideological positions.
In this setting, the \emph{share} of candidates of party $i$ with an ideological position between $a$ and $b$ is 
\ifbool{finalsubmission}
{$\mu_i((a,b)) = \int_a^b\d\mu_i(x),$}
{\begin{equation*}
\mu_i((a,b)) = \int_a^b\d\mu_i(x),
\end{equation*}}
and the party's average ideological position is
\ifbool{finalsubmission}
{$\mathbb{E}_{\mu_i}[y_i]=\int_{\mathbb{R}}y_i\d\mu_i(y_i).$}
{\begin{equation*}
    \mathbb{E}_{\mu_i}[y_i]=\int_{\mathbb{R}}y_i\d\mu_i(y_i).
\end{equation*}}

Accordingly, similarly to~\eqref{eq:background:Vtilde}, the expected total number of votes for party $1$ is
\begin{equation}\label{eq:V}
\begin{aligned}
    V_1(\mu_1,\mu_2)
    %&=\mathbb{E}_{\mu_2} \left[\mathbb{E}_{\mu_1}\left[\tilde{V_1}(y_1,y_2)]\right]\right]\\
    &=\int_\mathbb{R} \int_\mathbb{R} \tilde{V_1}(y_1,y_2)\d\mu_1(y_1)\d\mu_2(y_2)\\
    &= \int_\mathbb{R}\int_\mathbb{R}\int_\mathbb{R}
    p_1(x|y_1,y_2)\d\rho(x)\d\mu_1(y_1)\d\mu_2(y_2).
\end{aligned}
\end{equation}
In plain words, $\tilde{V_1}(y_1,y_2)\d\mu_1(y_1)\d\mu_2(y_1)$ is the number of votes that party 1 receives considering the candidates of party 1 with ideology $y_1$ and the candidates of party 2 with ideology $y_2$ (cf. Eq.~\eqref{eq:background:Vtilde}). Thus, the total number of votes results from integration over the ideology space of the two parties.
Since $\tilde V_i$ is non-negative, by Tonelli's theorem~\cite{rudin1987}, the order of integration in~\eqref{eq:V} does not matter.  
Again, the expression for $V_2$ is symmetric.

As in~\cref{sec:background}, we suppose that parties aim at maximizing the expected total number of votes, and we adopt a gradient flow approach to model the continuous-time evolution of the parties' full ideological distributions. We resort to the theory of Wasserstein gradient flows, also known as gradient flows in the Wasserstein space. The Wasserstein space is the space of probability distributions with finite second moment endowed with the Wasserstein distance, defined by
\begin{equation*}
    \wassersteinDistance{\mu}{\nu}
    ={\left(\min_{\gamma\in\setPlans{\mu}{\nu}}\int_{\mathbb{R}\times\mathbb{R}} \abs{x-y}^2\d\gamma(x,y)\right)}^{\frac{1}{2}},
\end{equation*}
where $\setPlans{\mu}{\nu}\subset\Pp{2}{\mathbb{R}\times\mathbb{R}}$ is the set of joint probability distributions (referred to as transport plans) with marginals $\mu_1$ and $\mu_2$. We refer to the set of minimizers $\setOptimalPlans{\mu}{\nu}\subset\setPlans{\mu}{\nu}$ as the set of optimal transport plans~\cite{Villani2007,Santambrogio2015}. Intuitively, the Wasserstein distance represents the minimum transportation cost to transport one distribution $\mu$ into another distribution $\nu$, when moving one unit of mass from $x$ to $y$ costs $\abs{x-y}^2$.

In this setting, the gradient flow equations read
\begin{equation}\label{eq:dynamic model}
\begin{aligned}
     \dot{\mu}_1(t)&=k\nabla_{\mu_1}{V_1}(\mu_1(t),\mu_2(t))\quad\mu_1(0)=\mu_{1,0}\\
     \dot{\mu}_2(t)&=k\nabla_{\mu_2}{V_2}(\mu_1(t),\mu_2(t))\quad\mu_2(0)=\mu_{2,0},
\end{aligned}
\end{equation}
where $\mu_{1,0},\mu_{2,0}\in\Pp{2}{\mathbb{R}}$ are given initial conditions and $k\in\reals_{>0}$ is a positive constant (determined from empirical data).
The expressions $\dot{\mu}$ and $\nabla_{\mu}{V}$ are to be intended in the sense of Wasserstein~\cite{Ambrosio2008}.
Namely, the ``time derivative'' $\dot{\mu}(t)$ is the tangent vector of an absolutely continuous (w.r.t. the Wasserstein distance) trajectory of probability measures $t\mapsto\mu(t)$; it can be identified with the velocity vector field $v:\reals\to\reals$ solving (in the sense of distributions) the continuity equation
\ifbool{finalsubmission}
{
$\dot\mu(t)+\nabla\cdot(v(t)\mu(t))=0.$
}
{
\begin{equation*}
    \dot\mu(t)+\nabla\cdot(v(t)\mu(t))=0.
\end{equation*}
}
We refer to~\cite[Chapter 8]{Ambrosio2008} for details.
Instead, the ``Wasserstein gradient'' of a function $V:\Pp{2}{\reals}\to\reals$ at $\mu\in\Pp{2}{\reals}$ is denoted by $\nabla_{\mu}{V}(\mu):\reals\to\reals$ and is the square integrable function (w.r.t. the measure $\mu$) which approximates $V$ ``linearly''; i.e., 
\begin{equation*}
    V(\nu)=\int_{\reals\times\reals} \left(\nabla_{\mu}{V}(\mu)(x)\right)(y-x)\d\gamma(x,y)+o(\wassersteinDistance{\mu}{\nu}).
\end{equation*}
for any optimal transport plan $\gamma\in\setOptimalPlans{\mu}{\nu}$ between $\mu$ and $\nu$. Here, $o(\wassersteinDistance{\mu}{\nu})$ denotes terms which are at least quadratic in the Wasserstein distance. For details, we refer to~\cite[Chapter 10]{Ambrosio2008} and~\cite{bonnet2019}. We will study the convergence properties of this model in the next section.

\subsection{Theoretic Analysis}\label{subsec:ourmodel:theory}

For our theoretic analysis, we assume that a sufficiently regular solution to~\eqref{eq:dynamic model} exists:

\begin{assumption}[Well-posed]\label{assumption:wellposed}
The dynamic system~\eqref{eq:dynamic model} admits a locally absolutely continuous solution $\mu_i:[0,+\infty)\to\Pp{2}{\reals}$ such that $\mu_i(0)=\mu_{i,0}$ for $i\in\{1,2\}$.
\end{assumption}

Note that we are \emph{not} assuming that $\mu_i(t)$ is absolutely continuous with respect to the Lebesgue measure, but that the curve $t\mapsto\mu_i(t)$ is absolutely continuous (seen as a curve between two metric spaces). \cref{assumption:wellposed} holds for gradient flows of the form $\dot\mu=-\nabla_{\mu}V$~\cite[Chapter 11]{Ambrosio2008}, provided that $V$ is sufficiently well behaved. Its study for systems of the form of~\eqref{eq:dynamic model} is left to future research.

We now give an explicit expression for the Wasserstein gradients. This will help us to study equilibria, but also to implement~\eqref{eq:dynamic model} numerically: 

\begin{lemma}[Wasserstein gradient]\label{lemma:wasserstein gradient}
The Wasserstein gradient of $V_1:\Pp{2}{\reals}\to\reals$ (with respect to $\mu_1$) at $(\mu_1,\mu_2)$ is the function
\ifbool{finalsubmission}{
$\nabla_{\mu_1}V_1(\mu_1,\mu_2):\reals \to \reals$ defined by 
\begin{equation*}
\begin{aligned}
    \nabla_{\mu_1}V_1(\mu_1,\mu_2)(y_1) = \int_{\reals}\nabla_{y_1}\tilde V_1(y_1,y_2)\d\mu_2(y_2),
\end{aligned}
\end{equation*}
}
{
\begin{equation*}
\begin{aligned}
    \nabla_{\mu_1}V_1(\mu_1,\mu_2):\reals &\to \reals \\
    y_1 & \mapsto \int_{\reals}\nabla_{y_1}\tilde V_1(y_1,y_2)\d\mu_2(y_2),
\end{aligned}
\end{equation*}
}
where $\nabla_{y_1}\tilde V_1(y_1,y_2)$ is the usual gradient of the real-valued function $\tilde V_1(y_1,y_2)$ defined in~\eqref{eq:background:Vtilde}.
The expression for $\nabla_{\mu_2}V_2$ is analogous.
\end{lemma}

We refer to the appendix for a proof. 
Armed with an explicit expression for Wasserstein gradients, we can now study the equilibria of~\eqref{eq:dynamic model}. As usual, $(\mu_1^\ast,\mu_2^\ast)$ is defined to be an equilibrium if the right hand side of~\eqref{eq:dynamic model} evaluates to 0.
Therefore, we look for $(\mu_1^\ast,\mu_2^\ast)$ so that the Wasserstein gradients $\nabla_{\mu_i}V_i(\mu_1,\mu_2)$ evaluate to the zero function in $L^2(\reals,\reals;\mu_i)$ (i.e., $\mu_i$-a.e.).
The resulting ``distributional'' equilibria are compatible with the point-wise equilibria of the satisficing model (cf. \cref{prop:background}):

\begin{lemma}[Equilibria]\label{lemma:equilibria}
Let~\cref{assumption:parameters} hold.
Then, the dynamic system~\eqref{eq:background:dynamic system} admits the following equilibria:
\begin{itemize}
    \item a symmetric equilibrium $\mu_1^\ast=\mu_2^\ast=\diracDelta{0}$;
    \item two asymmetric equilbria $\mu_1^\ast=\diracDelta{y^\ast}$, $\mu_2^\ast=\diracDelta{-y^\ast}$, where $y^\ast$ results from \eqref{eq:background:asymmetric equilbrium}.
\end{itemize}
Moreover, if~\cref{assumption:parameters} does not hold, then~\eqref{eq:background:dynamic system} admits the equilibrium $\mu_1^\ast=\mu_2^\ast=\diracDelta{0}$.
\end{lemma}

The proof is reported in the appendix.
\cref{lemma:equilibria} does not characterize all equilibria, but it suggests that some of the equilibria of~\eqref{eq:background:dynamic system} are delta distributions, namely ideological distributions where all candidates share the same ideological position.
In the next theorem, we show that some of these equilibria are attractive:

\begin{theorem}[Convergence]\label{thm:stability}
Let \cref{assumption:parameters} hold and let $y^\ast$ as in~\eqref{eq:background:asymmetric equilbrium}.
Then, there exists $\varepsilon>0$ such that if $\mu_{1,0}, \mu_{2,0}\in\Pp{2}{\reals}$ are supported on $y^\ast+[-\varepsilon,\varepsilon]$ and $-y^\ast+[-\varepsilon,\varepsilon]$, then $\mu_1(t)$ and $\mu_2(t)$ converge weakly to $\delta_{y^\ast}$ and $\delta_{-y^\ast}$ in $\Pp{2}{\reals}$\footnote{We say that $\mu(t)$ converges \emph{weakly in $\Pp{2}{\reals}$} to $\mu^\ast$ if for all continuous functions $f:\reals\to\reals$ with $|f(x)|\leq A+Bx^2$, $A,B\in\reals$, we have
%\begin{equation*}
    $\lim_{t\to\infty}\int_{\reals} f(y)\d\mu(t)(y)=\int_{\reals}f(y)\d\mu^\ast(y)$.
%\end{equation*}
}, respectively.
Similarly, if $\mu_{1,0}, \mu_{2,0}\in\Pp{2}{\reals}$ are supported on $-y^\ast+[-\varepsilon,\varepsilon]$ and $y^\ast+[-\varepsilon,\varepsilon]$, then $\mu_1(t)$ and $\mu_2(t)$ converge weakly to $\delta_{-y^\ast}$ and $\delta_{+y^\ast}$ in $\Pp{2}{\reals}$, respectively. 
Moreover, if~\cref{assumption:parameters} does not hold, then there exists $\varepsilon>0$ such that if $\mu_{1,0}, \mu_{2,0}\in\Pp{2}{\reals}$ are supported on $[-\varepsilon,\varepsilon]$, then $\mu_1(t)$ and $\mu_2(t)$ both converge weakly in $\Pp{2}{\reals}$ to $\delta_{0}$ in $\Pp{2}{\reals}$.
\end{theorem}

We refer to the appendix for a proof. 
In plain words, \cref{thm:stability} implies that whenever the support of the initial ideological distributions is sufficiently close to the equilibrium, then the ideological distributions of both parties converge (weakly) to two delta distributions, supported at the equilibrium of the satisficing model from~\cite{Yang2020}.
This allows for the following interpretation: Parties eventually become entirely homogeneous, with all candidates converging to the same ideological position. We will provide empirical evidence of the conclusions of~\cref{thm:stability} in the next section.

\begin{remark}
\cref{thm:stability} does not provide a notion of local asymptotic stability.
For instance, it does not allow us to conclude that $(\delta_{y^\ast},\delta_{-y^\ast})$ is locally asymptotically stable (with stability defined with respect to the Wasserstein distance). 
Indeed, for all $\eta>0$, there exists $\tilde\mu_1$ $\eta$-close to $\diracDelta{y^\ast}$ (i.e., $\wassersteinDistance{\diracDelta{y^\ast}}{\tilde\mu_1}\leq\eta$) not supported on $y^\ast+[-\varepsilon,\varepsilon]$ (and thus for which \cref{thm:stability} does not apply); e.g., for all $n\in\naturals$ with $n$ sufficiently large
\ifbool{finalsubmission}{
$\tilde\mu_1\coloneqq(1-\frac{\eta^2}{n^2})\delta_{y^\ast}+\frac{\eta^2}{n^2}\delta_{y^\ast-n}$
}
{
\begin{equation*}
    \tilde\mu_1\coloneqq\left(1-\frac{\eta^2}{n^2}\right)\delta_{y^\ast}+\frac{\eta^2}{n^2}\delta_{y^\ast-n}
\end{equation*}
}
is $\eta$-close to $\diracDelta{y^\ast}$, since
\ifbool{finalsubmission}{
$\wassersteinDistance{\tilde\mu_1}{\diracDelta{y^\ast}}=\sqrt{(1-\frac{\eta^2}{n^2})\cdot 0 + \frac{\eta^2}{n^2}n^2}=\eta,$
}
{
\begin{equation*}
    \wassersteinDistance{\tilde\mu_1}{\diracDelta{y^\ast}}
    =
    \sqrt{\left(1-\frac{\eta^2}{n^2}\right)\cdot 0 + \frac{\eta^2}{n^2}n^2}=\eta,
\end{equation*}
}
but it is clearly \emph{not} supported on $y^\ast+[-\varepsilon,\varepsilon]$. We leave the study of local asymptotic stability region to future work.
\end{remark}

\subsection{Discussion}\label{subsec:ourmodel:discussion}
Few comments are in order. 
First, we do \emph{not} restrict ourselves to a specific class of probability distributions (e.g., Gaussian, continuous, or discrete), but we work in the probability space $\Pp{2}{\reals}$, which includes \emph{all} probability distributions over the real line, provided that their second moment is finite.
Second, since probability distributions are normalized, $\mu_i((a,b))$ is \emph{not} the total number of candidates with an ideological position between $a$ and $b$, but the share of candidates. This way we can directly deploy the rich theory of optimal transport, formalized for probability distributions, without introducing normalization terms. 
%Third, if $\mu_i$ is absolutely continuous with respect to the Lebesgue measure, then it has a probability density (say, with a slight abuse of notation, $\mu_i(y_i))$, and we can replace all $\d\mu_i(y_i)$ with the more standard $\mu_i(y_i)\d y_i$.
Third, we do not prove that $\mu_i(t)(A)$ converges to $\mu_i^\ast(A)$ for all Borel sets $A$ (i.e., \emph{strong} convergence), but that the integral of each continuous function with quadratic growth converges (i.e., \emph{weak} convergence). The interpretation is as follows: We do not perform a ``microscopic'' analysis on each portion of the ideology space, but rather a ``macroscopic'' analysis for all aggregated quantities resulting from an integral (such as mean, number of votes, second moment, etc.).
Fourth, our model predicts convergence to delta distributions, representing homogeneous parties. Yet, it can be regularized (e.g., via an entropy term), so that equilibria yield more heterogeneous ideologies. We leave this analysis to future work.

\section{Results}\label{sec:results}
In this section, we present our numerical results. We present a simulation in~\cref{subsect:simulations} and compare our model with data in \cref{subsect:validation}. In the appendix, we study the setting with three parties.

\subsection{Simulations}\label{subsect:simulations}
For simulation purposes, we approximate all probability distributions (i.e., ideological distributions) with discrete measures of 300 particles (representing 300 candidates) and approximate the dynamics~\eqref{eq:dynamic model} by
\begin{equation}\label{eq:implementation}
\begin{aligned}
    \mu_1(t+1)&=\pushforward{(\Id + \tau k \nabla_{\mu_1}V_1(\mu_1(t),\mu_2(t))}\mu_1(t)
    \\
    \mu_2(t+1)&=\pushforward{(\Id + \tau k \nabla_{\mu_2}V_2(\mu_1(t),\mu_2(t))}\mu_2(t),
\end{aligned}
\end{equation}
where $\Id$ is the identity map on $\reals$, $\tau\in\reals_{>0}$ is the step-size, and $\pushforward{(\cdot)}$ denotes the pushforward operator for probability measures~\cite{Ambrosio2008}. In turn, \eqref{eq:implementation} stipulates that a particle of $\mu_1$ at position $x$ is displaced to  $x + \tau k\nabla_{\mu_1}V_1(\mu_1(t),\mu_2(t))(x)$.
The public's distribution $\rho(x)$ is a zero-mean Gaussian distribution with standard deviation $\sigma_0=0.93$, determined from data of the US Congress~\cite{Yang2020}.
We use the nominal parameters $k=0.5$ and $\sigma=0.6$.
The initial distributions $\mu_{1,0}$ and $\mu_{2,0}$ are samples from truncated Gaussian distributions (truncated at $0$ and $0.8$, and $-0.8$ and $0$, respectively), originally with mean $-0.25$ (for party 1), $0.25$ (for party 2), and standard deviation $0.15$ (for both parties); see~\cref{fig:ICsim}.
\ifbool{finalsubmission}{
%\begin{figure}[t] % arxiv
\begin{figure}[b] % final submission
    \centering
    \includegraphics[width=\widthfigures]{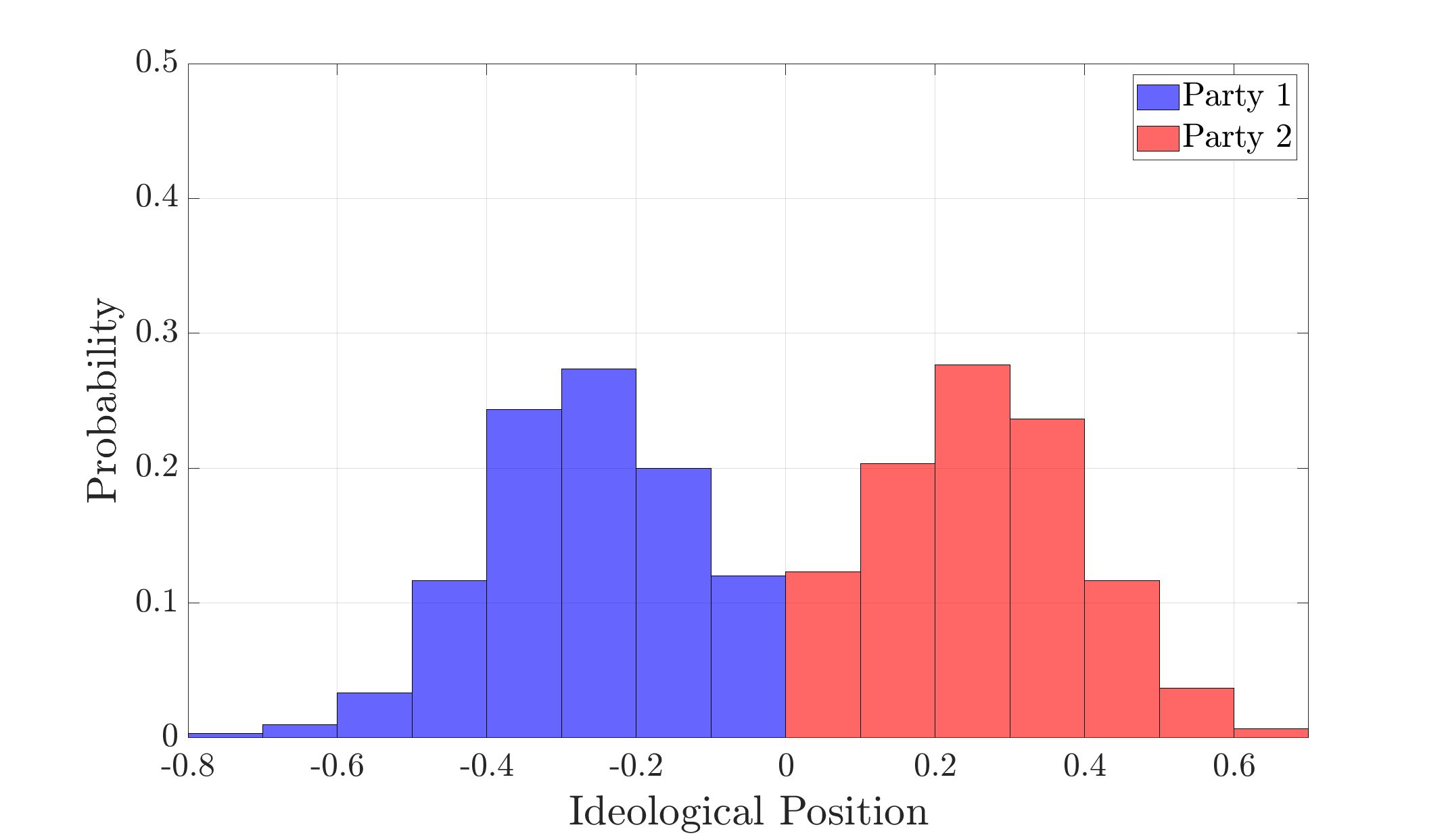}
    \caption{Initial ideological distributions $\mu_{1,0}$ and $\mu_{2,0}$, defined as samples from truncated Gaussian distributions (truncated at $0$ and $0.8$, and $-0.8$ and $0$, respectively), originally with mean $\pm 0.25$ and standard deviation $0.15$.}
    \label{fig:ICsim}
\end{figure}
}{
\begin{figure}[t] % arxiv
%\begin{figure}[b] % final submission
    \centering
    \includegraphics[width=\widthfigures]{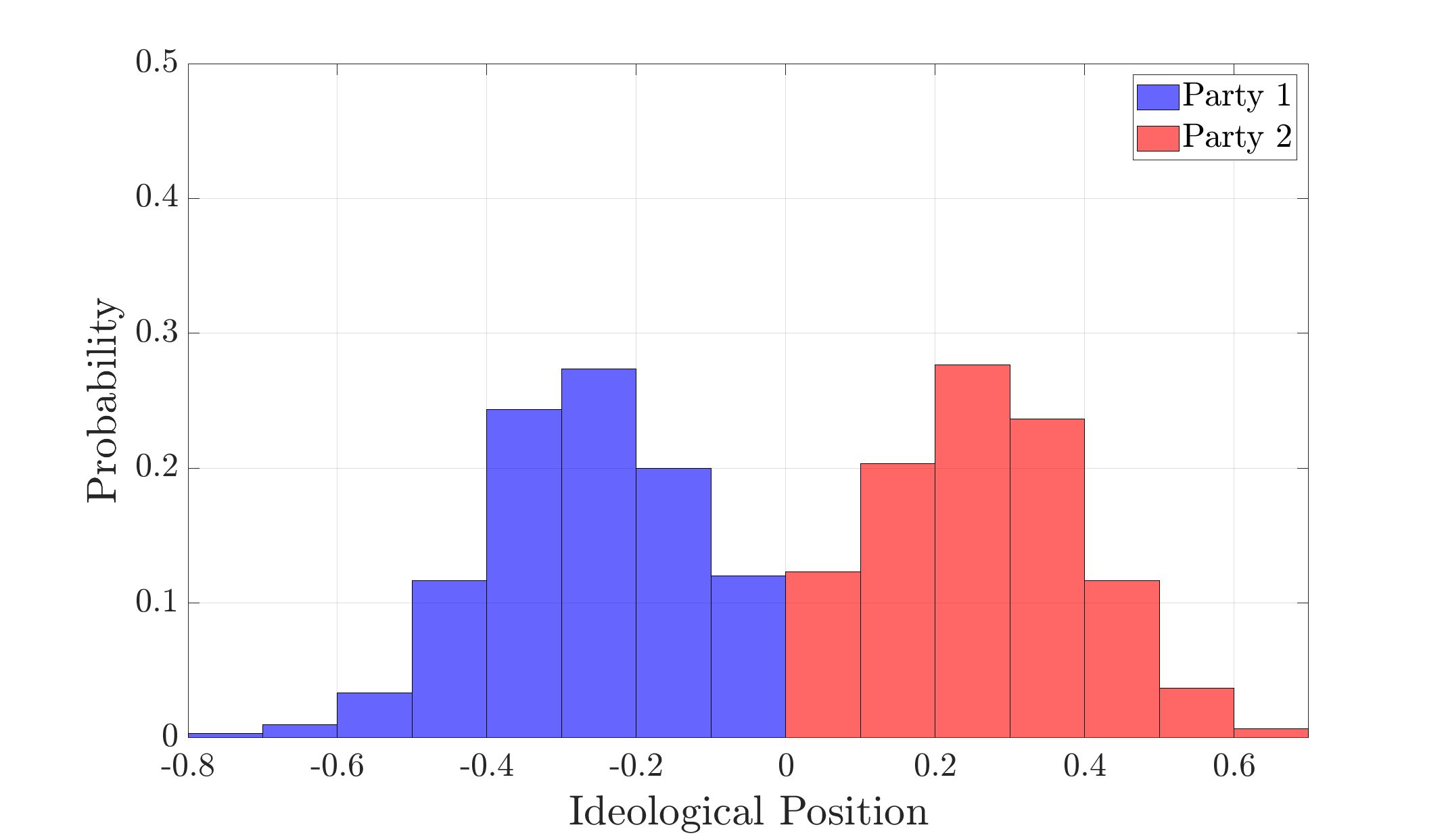}
    \caption{Initial ideological distributions $\mu_{1,0}$ and $\mu_{2,0}$, defined as samples from truncated Gaussian distributions (truncated at $0$ and $0.8$, and $-0.8$ and $0$, respectively), originally with mean $\pm 0.25$ and standard deviation $0.15$.}
    \label{fig:ICsim}
\end{figure}
}

\begin{figure}[t]
   \begin{center}
    \includegraphics[width=\widthfigures]{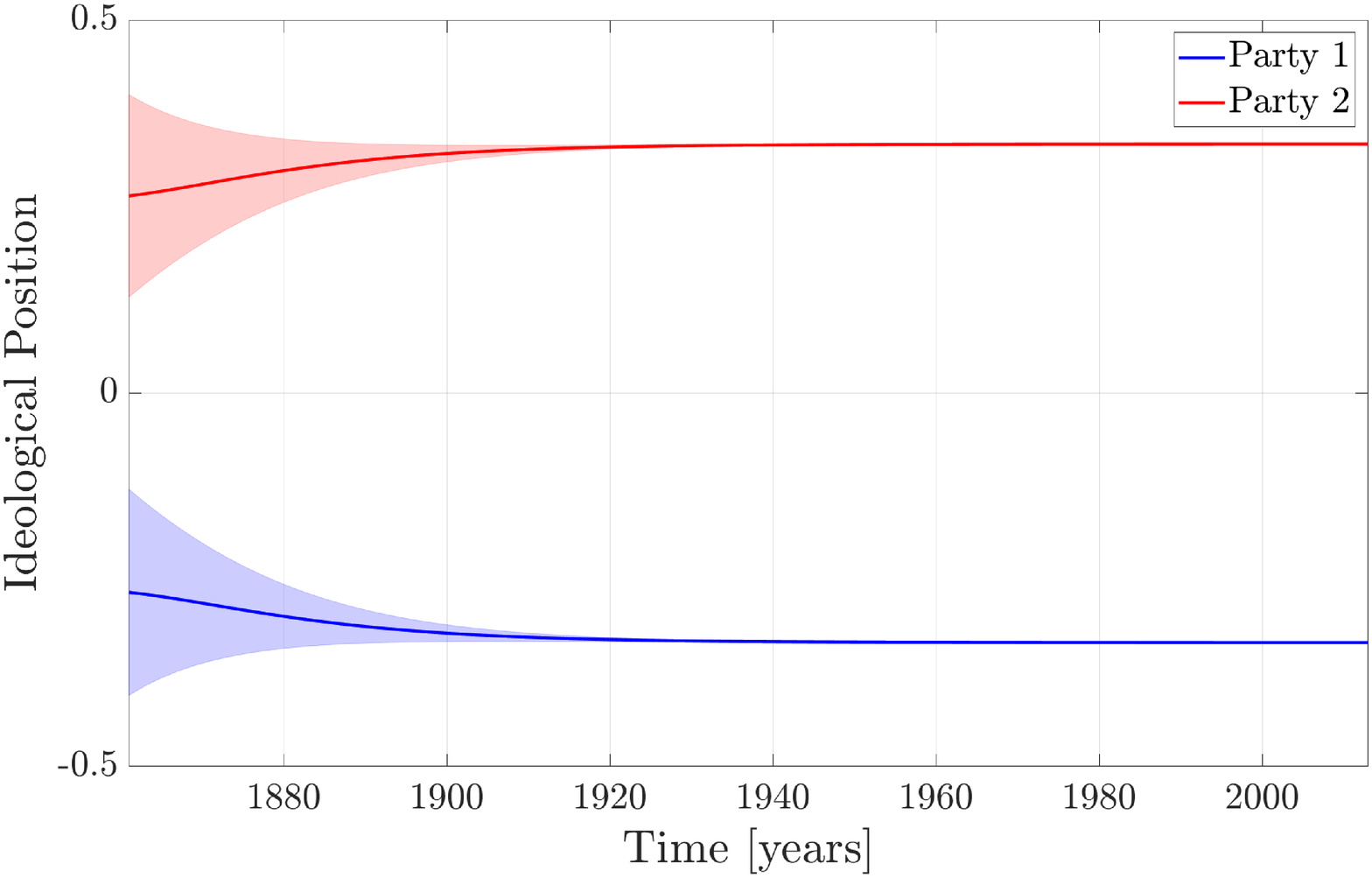}
    \caption{Time evolution of the parties' ideological distributions (mean plus/minus standard deviation), with $k=0.5$ and $\sigma=0.6$. The initial distributions result from samples of truncated Gaussian distributions; see~\cref{fig:ICsim}. The experiments confirm the theoretic results: Parties' polarization increases, and parties becomes more and more homogeneous.}
    \label{fig:polc1b}
    \end{center}
    \ifbool{finalsubmission}{\vspace{-0.4cm}}{}
\end{figure}

\begin{figure}[t]
   \begin{center}
    \includegraphics[width=\widthfigures]{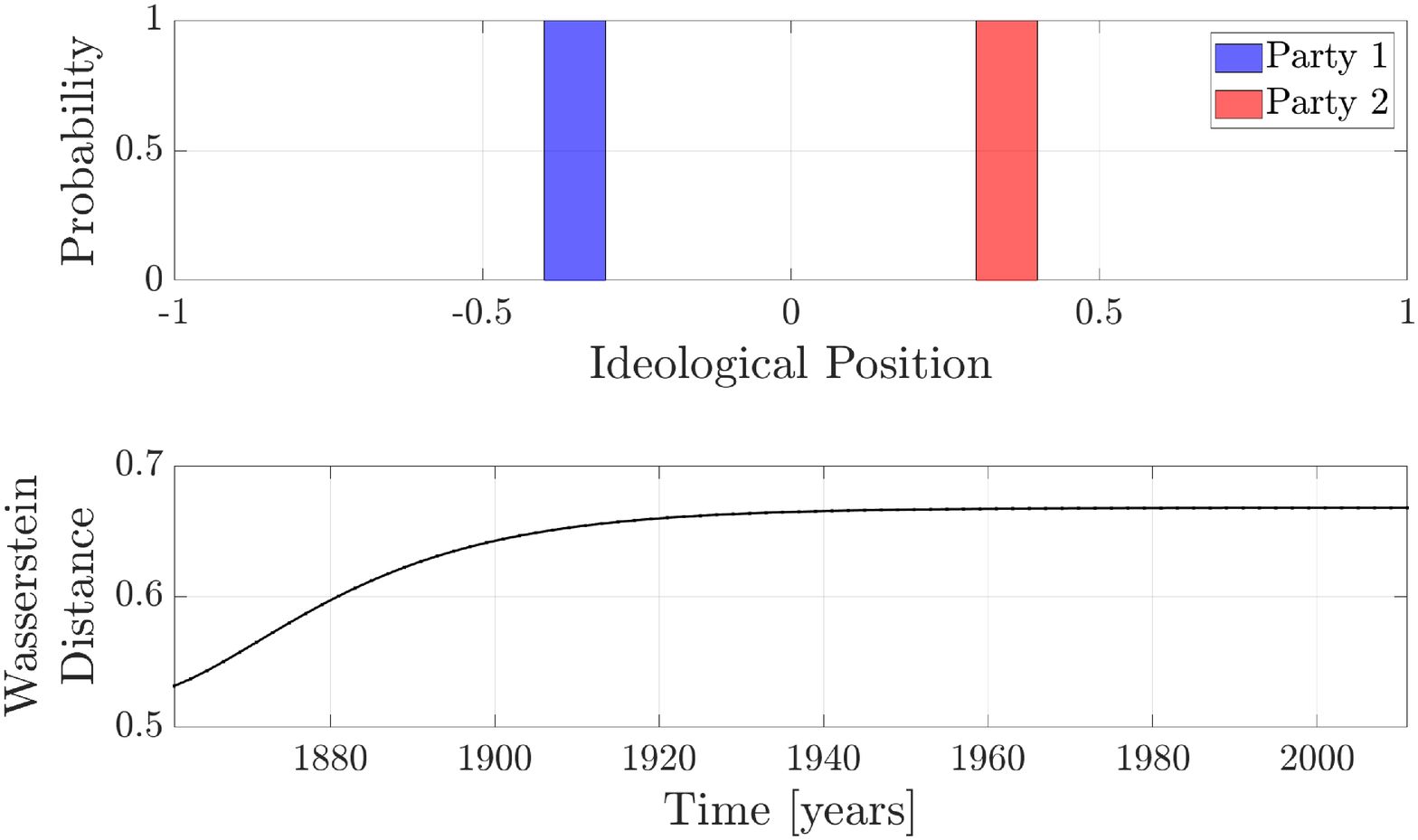}
    \caption{Top: ideological distributions of the two parties at the end of the horizon. Parties' ideologies converge to delta distributions, showing that parties eventually become homogeneous. Bottom: time evolution of the Wasserstein distance between the parties' ideological distributions, showing that polarization increases monotonically.}
    \label{fig:polc1a}
    \end{center}
    \ifbool{finalsubmission}{\vspace{-0.4cm}}{}
\end{figure}

We show the results of our simulations in~\cref{fig:polc1a,fig:polc1b}.
As can be seen in~\cref{fig:polc1b}, our approach is indeed capable of modeling the time evolution of the parties' ideological distributions. Thus, we can infer features such as parties' inclusiveness and homogeneity, and not only their average ideological positions. 
Our simulation confirms that parties become more polarized with time and less inclusive (\cref{fig:polc1b}), until they both converge to two distinct delta distributions (\cref{fig:polc1a}, top), as predicted by~\cref{thm:stability}.
Again, this result allows for the following interpretation: Parties eventually become homogeneous, with all candidates sharing the same ideological position. 
At equilibrium, parties get the same number of votes, and 27\% of the public does not vote.
Finally, our model predicts that political polarization increases monotonically with time and eventually converges: The Wasserstein distance (\cref{fig:polc1a}, bottom) between the ideological distributions increases monotonically, and converges to $0.67$.


\subsection{Validation and Parameters Fitting}\label{subsect:validation}
\begin{figure}[t]
   \begin{center}
    \includegraphics[width=\widthfigures]{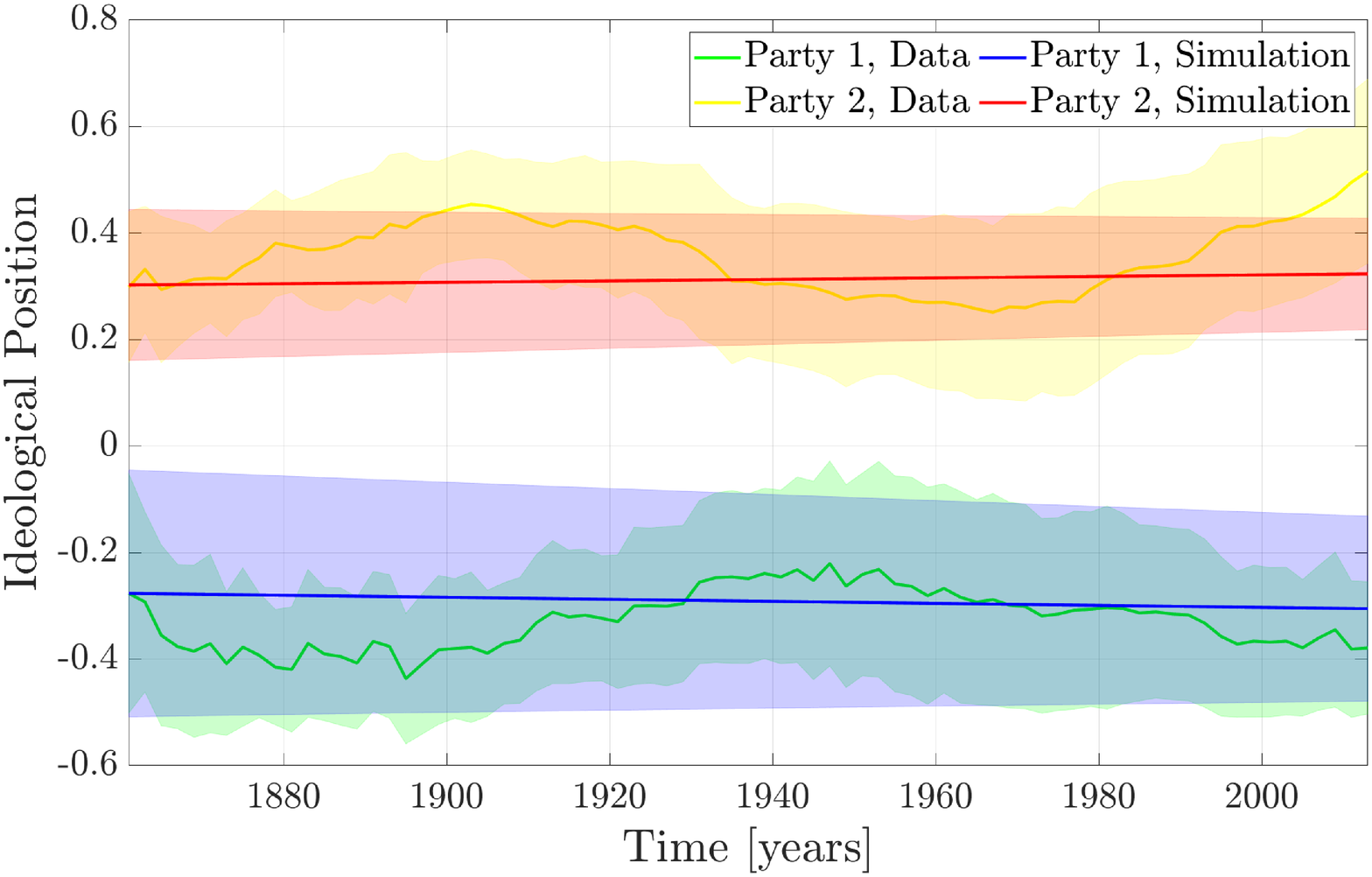}
    \caption{Comparison of our model with data. For all ideologies we plot average plus/minus standard deviation. In green and yellow: ideological distribution of each party, using data from~\cite{dataa}. In red and blue: predictions of our model, with the fitted parameters (i.e., $k=0.026$, $\sigma=0.389$) and initial (samples of) Gaussian distributions inferred from data of 1861.}
    \label{fig:fit}
    \end{center}
\end{figure}

We validate our model with data from the US Congress, as in~\cite{Yang2020}.
Specifically, we use a combined dataset of representatives and senators of the Democratic and Republican parties in the US Congress~\cite{dataa}. The dataset comprises the ideology score of every candidate in each party, during the period 1861--2015. \cref{fig:original} shows the time evolution of the parties' average ideological position, together with their standard deviations.
We fit the parameters $k$ and $\sigma$ to minimize the mean squared error, quantified via the Wasserstein distance between the true and the predicted ideological distribution of every party. Formally, given the true trajectory $\{(\hat\mu_1(0),\hat\mu_2(0)),\ldots, (\hat\mu_1(T),\hat\mu_2(T))\}$ for a horizon of length $T$, we solve
\begin{equation*}
\begin{aligned}
    k,\sigma \in \argmin
    &\frac{1}{T}\sum_{t=1}^T\wassersteinDistance{\hat\mu_1(t)}{\mu_1(t)}^2+\wassersteinDistance{\hat\mu_2(t)}{\mu_2(t)}^2
    \\
    &\text{s.t. } \eqref{eq:implementation}, \mu_1(0)=\hat\mu_1(0), \mu_2(0)=\hat\mu_2(0).
\end{aligned}
\end{equation*}
The identification yields $k=0.0264$ and $\sigma=0.389$.
The comparison of the model's performance with data is shown in~\cref{fig:fit}. Our model captures the overall behavior of the data. However, it disregards oscillations that are probably due to exogenous impact factors, such as the historical context, election rounds, and political campaigns.

\section{Conclusion}\label{sec:conclusion}
We presented a satisficing dynamical model for political competition between two parties. Rather than lumping parties in their average ideological position as in~\cite{Yang2020}, our model predicts the dynamic behavior of the full ideological distribution.
Under the assumption that parties aim at maximizing the expected total number of votes, we formulated a Wasserstein gradient flow for the time evolution of their ideological distributions.
Our model predicts that parties become more homogeneous and polarized with time, until their ideological distributions converge to asymmetric delta distributions.
We provided theoretic and numerical support for our findings, and we validated our model with data from the US Congress.

Our model captures the trend in the data, but it disregards impact factors such as the historical context, election rounds, and political campaigns. These aspects, together with further theoretic analysis (e.g., regularization), connections with dynamic game theory~\cite{bacsar1998dynamic} and uncertainty propagation~\cite{aolaritei2022}, and case studies (e.g., asymmetric initial ideological distributions), are possible avenues for future research. 
%%%%%%%%%%%%%%%%%%%%%%%%%%%%%%%%%%%%%%%%%%%%%%%%%%%%%%%%%%%%%%%%%%%%%%%%%%%%%%%%
\ifbool{finalsubmission}{}{
\appendix
\subsection{Proofs}
\begin{proof}[Proof of~\cref{prop:background}]
The proof follows directly from~\cite[Section 3]{Yang2020}. Local asymptotic stability follows via the linearization method~\cite{Khalil2002}.
\end{proof}

\begin{proof}[Proof of~\cref{lemma:wasserstein gradient}]
First, we rewrite $V_1$ as 
\begin{equation*}
    V_1(\mu_1,\mu_2)
    =
    \int_\mathbb{R} \left(\int_\mathbb{R} \tilde{V_1}(y_1,y_2)\d\mu_2(y_2)\right)\d\mu_1(y_1).
\end{equation*}
Then, by~\cite[Proposition 10.4.2., Remark 10.4.3]{Ambrosio2008}, we have 
\begin{equation*}
\begin{aligned}
    \nabla_{\mu_1}V_1(\mu_1,\mu_2)
    &=
    \nabla_{y_1}\left(\int_\mathbb{R} \tilde{V_1}(y_1,y_2)\d\mu_2(y_2)\right)
    \\
    &=
    \int_\mathbb{R} \nabla_{y_1}\tilde{V_1}(y_1,y_2)\d\mu_2(y_2),
\end{aligned}
\end{equation*}
where the last equality follows from dominated convergence since $\tilde{V_1}(y_1,y_2)$ is continuously differentiable with bounded derivatives~\cite{rudin1987}.
\end{proof}

\begin{proof}[Proof of~\cref{lemma:equilibria}]
We only prove the statement when~\cref{assumption:parameters} holds true; the other case follows \emph{mutatis mutandis}. We seek to prove that at $(\diracDelta{y^\ast},\diracDelta{-y^\ast})$ the Wasserstein gradients coincide with the zero function.

Let $\mu_i$ be delta distributions; i.e., $\mu_i=\delta_{\bar y_i}$ for some $\bar y_i\in\reals$. Then, the Wasserstein gradients are 
\begin{equation*}
\begin{aligned}
    \nabla_{\mu_1}V_1(\mu_1,\mu_2)&=\nabla_{y_1}\tilde V_1(y_1,\bar y_2)=\nabla_{y_1}\tilde V_1(\bar y_1,\bar y_2)\quad \mu_1\text{-a.e.},
    \\
    \nabla_{\mu_1}V_2(\mu_1,\mu_2)&=\nabla_{y_1}\tilde V_2(\bar y_1,y_2)=\nabla_{y_2}\tilde V_2(\bar y_1,\bar y_2)\quad \mu_2\text{-a.e.},
\end{aligned}
\end{equation*}
which evaluate to zero $\mu_i$-a.e. if and only if $\nabla_{y_i}\tilde V_i(\bar y_1,\bar y_2)=0$.
From~\cref{prop:background}, $\nabla_{y_1}\tilde V_1(\bar y_1,\bar y_2)$ and $\nabla_{y_2}\tilde V_2(\bar y_1,\bar y_2)$ are simultaneously zero if and only if $\bar y_1=\bar y_2=0$ or $\bar y_1=-\bar y_2=y^\ast$ with $y^\ast$ as in~\eqref{eq:background:asymmetric equilbrium}. Thus, $\mu_1=\mu_2=\delta_0$, and $\mu_1=\delta_{y^\ast},\mu_2=\delta_{-y^\ast}$ are equilibria. In fact, the proof also reveals that they are the only equilibria where both $\mu_1$ and $\mu_2$ are delta distributions. 
\end{proof}

\begin{proof}[Proof of~\cref{thm:stability}]
We only prove the statement when~\cref{assumption:parameters} holds true; the other cases follow \emph{mutatis mutandis}. The proof proceeds in three steps.
First, we prove that there exists $\varepsilon>0$ and $\alpha>0$ such that for all $(y_1,y_2)\in K\coloneqq K_1\times K_2$ with $K_1\coloneqq y^\ast+[-\varepsilon,\varepsilon]$ and $K_2=-y^\ast+[-\varepsilon,\varepsilon]$:
\begin{equation}\label{eq:proof:ineq}
\begin{aligned}
    \nabla_{y_1}\tilde V_1(y_1,y_2)&(y_1-y^\ast) + \nabla_{y_2}\tilde V_2(y_1,y_2)(y_2+y^\ast) \\
    &\leq
    -\alpha\left((y_1-y^\ast)^2+(y_2+y^\ast)^2\right).
\end{aligned}
\end{equation}
Second, we prove that if $\mu_{i,0}$ is supported on $K_i$, then $\mu_i(t)$ is supported on $K_i$ for all times. 
Third, we compute the time derivative of the Wasserstein distance between $\mu_i(t)$ and equilibrium $\mu^\ast$, and show that it converges to 0.

To prove~\eqref{eq:proof:ineq}, define $P:\reals\times\reals\to\reals$ by
\begin{equation*}
    P(y_1,y_2)=\int_{\reals}s_1(d_1)+s_2(d_2)-\frac{1}{2}s_1(d_1)s_2(d_2)\d\rho(x),
\end{equation*}
where $d_i=|x-y_i|$. By construction, $\nabla_{y_i}P(y_1,y_2)=\nabla_{y_i}\tilde V_i(y_1,y_2)$ for $i\in\{1,2\}$.
Moreover, $P$ is strongly concave in a neighborhood $U\subset\reals^2$ of $(y^\ast,-y^\ast)$, and locally maximized at $(y^\ast,-y^\ast)$; else, the equilibrium $(y^\ast,-y^\ast)$ would not be locally asymptotically stable of the background model (cf.~\cref{prop:background}). Define $\varepsilon>0$ so that $K\subset U$; note that $\varepsilon$ is well-defined by definition of product topology.
By local strong concavity, there exists $\alpha>0$ (also called concavity parameter) such that for all $(y_1, y_2)\in K$ we have 
\begin{equation*}
\begin{aligned}
    P(y^\ast,-y^\ast)
    &\leq P(y_1,y_2)+\nabla_{y_1} P(y_1,y_2)(y^\ast-y_1)
    \\
    &+\nabla_{y_2} P(y_1,y_2)(-y^\ast-y_2)
    \\
    &-\alpha\left((y_1-y^\ast)^2+(y_2+y^\ast)^2\right).
\end{aligned}
\end{equation*}
Basic algebraic manipulations, together with local optimality of $(y^\ast,-y^\ast)$, lead to~\eqref{eq:proof:ineq}.

We now prove that if $\mu_{1,0}$ is supported on $K_1$ and $\mu_{2,0}$ is supported on $K_2$, then $\mu_1(t)$ and $\mu_2(t)$ are supported on $K_1$ and $K_2$ for all times. 
Observe that, possibly up to choosing a smaller $\varepsilon>0$, the vector
\begin{equation*}
    \left(k\int_{\reals}\nabla_{y_1}\tilde V_1(y_1,y_2)\d\mu_2(y_2),
    k\int_{\reals}\nabla_{y_2}\tilde V_2(y_1,y_2)\d\mu_1(y_1)\right),
\end{equation*}
points inwards for all $(y_1,y_2)\in K$, being the expected value of inward pointing vectors. For instance, at $y^\ast+\varepsilon$, we have that $k\nabla_{y_1}\tilde V_1(y^\ast+\varepsilon,y_2)<0$ for all $y_2\in K_2$, and so $\int_{\reals}\nabla_{y_1}\tilde V_1(y^\ast+\varepsilon,y_2)\d\mu_2(y_2)<0$.
Consider now a compactly supported smooth function $\xi:\reals\to\reals$ such that $\xi(x)=1$ on $K_1$, $\xi(x)<1$ outside of $K_1$, and $\xi(x)=0$ outside an open set containing $K_1$. Then, since the curve $t\mapsto\mu_1(t)$ is absolutely continuous, \cite[Theorem 8.3.1]{Ambrosio2008} shows that it solves the continuity equation in the sense of distributions:
\begin{equation*}
\begin{aligned}
    &\frac{\d}{\d t}\int \zeta(y_1,t)\d\mu_1(t)(y_1)
    \\
    &=\int_{\reals}\int_{\reals}  \nabla_{y_1}\zeta(y_1,t)\nabla_{\mu_1}V_1(y_1,y_2)\d\mu_1(t)(y_1)
    \\
    &=k\int_{\reals}\int_{\reals}  \nabla_{y_1}\zeta(y_1,t)\nabla_{y_1}\tilde V_1(y_1,y_2)\d\mu_2(t)(y_2)\d\mu_1(t)(y_1)
\end{aligned}
\end{equation*}
for all $\zeta:(a,b)\times\reals\to\reals$ smooth and with compact support, with $a<b$. We can now choose $\zeta(t,x)=\xi(x)$ to get
\begin{equation*}
\begin{aligned}
    &\frac{\d}{\d t}\int \xi(y_1)\d\mu_1(t)(y_1)
    \\
    &=k\int_{\reals}\int_\reals  \nabla_{y_1}\xi(y_1)\nabla_{y_1}\tilde V_1(y_1,y_2)\d\mu_2(t)(y_2)\d\mu_1(t)(y_1)
    \\
    &\geq 0,
\end{aligned}
\end{equation*}
since $\nabla_{y_1}\xi(y_1)\nabla_{y_1}\tilde V_1(y_1,y_2)$ is always non-negative (in particular, $\nabla_{y_1}\xi(y_1)$ and $\nabla_{y_1}\tilde V_1(y_1,y_2)$ have the same sign). 
Since $\mu_{1,0}$ is supported on $K_1$, we have $\int_{\reals}\xi(y_1)\d\mu_1(0)(y_1)=1$, and so $\int_{\reals}\xi(y_1)\d\mu_1(t)(y_1)=1$ for all $t\in\reals_{>0}$. Thus, $\mu_1(t)$ is supported on $K_1$ for all times. Analogously, we conclude that $\mu_2(t)$ is supported on $K_2$ for all times. 

We now prove that the Wasserstein distance converges to 0. We can now leverage~\cite[Theorem 8.4.7]{Ambrosio2008} to compute the time derivative of the Wasserstein distance between $\mu_1(t)$ and $\delta_{y^\ast}$:
\begin{equation*}
\begin{aligned}
    &\frac{\d}{\d t}\wassersteinDistance{\mu_1(t)}{\delta_{y^\ast}}^2
    \\
    &=
    \int_{\reals}\nabla_{\mu_1}V_1(\mu_1,\mu_2)(y_1)(y_1-y^\ast)\d\gamma(t)(y_1,\bar y_1),
\end{aligned}
\end{equation*}
for almost all times and where $\gamma(t)\in\Gamma_o(\mu(t),\delta_{y^\ast})$. Since $T(y_1)=y^\ast$ is the unique optimal transport map from $\mu_1(t)$ to $\diracDelta{y^\ast}$ we have
\begin{equation}\label{eq:proof:wassersteinone}
\begin{aligned}
    &\frac{\d}{\d t}\wassersteinDistance{\mu_1(t)}{\delta_{y^\ast}}^2
    \\
    &=
    \int_{\reals}\nabla_{\mu_1}V_1(\mu_1,\mu_2)(y_1)(y_1-y^\ast)\d\mu_1(t)(y_1),
    \\
    &=
    \int_{\reals}\left(k\int_{\reals}\nabla_{y_1}\tilde V_1(y_1,y_2)\d\mu_2(t)(y_2)\right)(y_1-y^\ast)\d\mu_1(t)(y_1).
\end{aligned}
\end{equation}
Analogously, for almost all times
\begin{equation}\label{eq:proof:wassersteintwo}
\begin{aligned}
    &\frac{\d}{\d t}\wassersteinDistance{\mu_2(t)}{\delta_{-y^\ast}}^2
    \\
    &=
    \int_{\reals}\left(k\int_{\reals}\nabla_{y_2}\tilde V_2(y_1,y_2)\d\mu_1(t)(y_1)\right)(y_2+y^\ast)\d\mu_2(t)(y_2).
\end{aligned}
\end{equation}
Since $\mu_1(t)$ is supported on $K_1$ and $\mu_2(t)$ is supported on $K_2$, we combine \eqref{eq:proof:wassersteinone} and \eqref{eq:proof:wassersteintwo} with \eqref{eq:proof:ineq}
\begin{equation*}
\begin{aligned}
    &\frac{\d}{\d t}\left(\wassersteinDistance{\mu_1(t)}{\delta_{y^\ast}}^2
    +
    \wassersteinDistance{\mu_2(t)}{\delta_{-y^\ast}}^2\right)
    \\
    &\leq 
    -\alpha k\int_{\reals}\int_{\reals}(y_1-y^\ast)^2+(y_2+y^\ast)^2\d\mu_2(t)(y_2)\d\mu_1(t)(y_1)
    \\
    &=
    -\alpha k \left(\wassersteinDistance{\mu_1(t)}{\diracDelta{y^\ast}}^2+\wassersteinDistance{\mu_2(t)}{\diracDelta{-y^\ast}}^2\right),
\end{aligned}
\end{equation*}
where we leverage Fubini's theorem to change the order of integration~\cite{rudin1987}. As above, we used that $T(y_1)=y^\ast$ is the unique optimal transport map from $\mu_1(t)$ to $\diracDelta{y^\ast}$ and $T(y_2)=-y^\ast$ is the unique optimal transport map from $\mu_2(t)$ to $\diracDelta{-y^\ast}$.
We can now deploy the (integral) Gronwall lemma to conclude that $\lim_{t\to\infty}\wassersteinDistance{\mu_1(t)}{\diracDelta{y^\ast}}^2+\wassersteinDistance{\mu_2(t)}{\diracDelta{-y^\ast}}^2=0$, which directly yields $\lim_{t\to\infty}\wassersteinDistance{\mu_1(t)}{\diracDelta{y^\ast}}=0$ and $\lim_{t\to\infty}\wassersteinDistance{\mu_2(t)}{\diracDelta{-y^\ast}}=0$.
Since the Wasserstein distance metrizes weak convergence in $\Pp{2}{\reals}$~\cite[Theorem 6.9]{Villani2007}, we directly establish weak convergence in~$\Pp{2}{\reals}$.
\end{proof}
\subsection{Additional Result: Growth of an Emerging Party}\label{sec:results:growth}
We now consider the case of three competing parties. In this case, the probability that a voter at position $x$ is satisfied with party 1 reads (cf.~\cref{eq:background:probability})
\begin{equation*}
\begin{aligned}
    p_1(x|y_1,y_2,y_3) &=s_1(d_1)(1-s_2(d_2))(1-s_3(d_3)) \\ &+\frac{1}{3}s_1(d_1)s_2(d_2)s_3(d_3).
\end{aligned}
\end{equation*}
The expressions for parties 2 and 3 are symmetric.
Then, \cref{eq:V,eq:dynamic model} are updated accordingly.

We run simulation as in~\cref{subsect:simulations}, and use the same parameters. The initial distributions $\mu_{1,0}$, and $\mu_{2,0}$ are the same as shown in~\cref{fig:ICsim}, while $\mu_{3,0}$ is sampled from a Gaussian distribution with mean $0$ and standard deviation $0.15$.
As can be seen in~\cref{fig:three_mean}, the presence of a third party induces more polarization: The Wasserstein distance between the ideologies of party 1 and 2 converges to $1.78$, and the Wasserstein distance between the ideologies of party 2 and 3 converges to $0.89$, both larger than the $0.67$ observed in the setting with two parties. As a consequence, 43\% of the public refrains from voting, corresponding to an increase of 16\% compared to the setting of two parties. 
The third party remains centrist, but its heterogeneity increases for the first few years, before decreasing and eventually becoming entirely homogeneous. This is in contrast with the other two parties and with previous experiments, which yield monotonically decreasing heterogeneity. 

\begin{figure}[b]
    \vspace{-0.5cm}
    \begin{center}
    \includegraphics[width=\widthfigures]{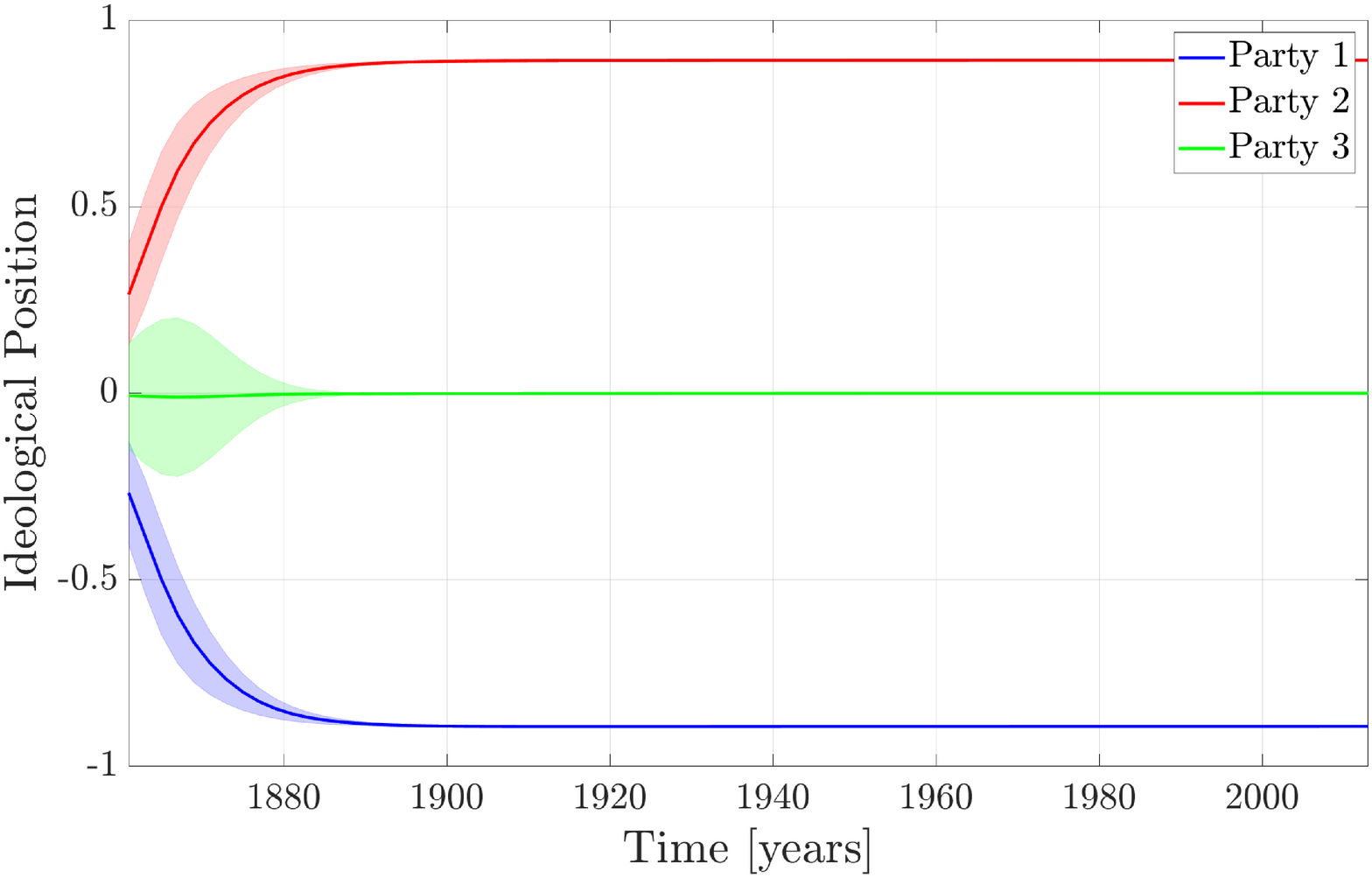}
    \caption{Time evolution of the parties' ideological distributions (mean plus/minus standard deviation), with $k=0.5$, $\sigma=0.6$, $\mu_{1,0}$ and $\mu_{2,0}$ as in~\cref{fig:ICsim}, and $\mu_{3,0}$ sampled from a Gaussian distribution with mean $0$ and standard deviation $0.15$. The experiment yields larger parties' polarization and initially increasing heterogenity for party 3.}
    \label{fig:three_mean}
    \end{center}
\end{figure}
}

%\IEEEtriggeratref{7}

\bibliographystyle{IEEEtran}
\bibliography{references}

\begin{thebibliography}{10}
\providecommand{\url}[1]{#1}
\csname url@samestyle\endcsname
\providecommand{\newblock}{\relax}
\providecommand{\bibinfo}[2]{#2}
\providecommand{\BIBentrySTDinterwordspacing}{\spaceskip=0pt\relax}
\providecommand{\BIBentryALTinterwordstretchfactor}{4}
\providecommand{\BIBentryALTinterwordspacing}{\spaceskip=\fontdimen2\font plus
\BIBentryALTinterwordstretchfactor\fontdimen3\font minus
  \fontdimen4\font\relax}
\providecommand{\BIBforeignlanguage}[2]{{%
\expandafter\ifx\csname l@#1\endcsname\relax
\typeout{** WARNING: IEEEtran.bst: No hyphenation pattern has been}%
\typeout{** loaded for the language `#1'. Using the pattern for}%
\typeout{** the default language instead.}%
\else
\language=\csname l@#1\endcsname
\fi
#2}}
\providecommand{\BIBdecl}{\relax}
\BIBdecl

\bibitem{fiorina2008political}
M.~P. Fiorina and S.~J. Abrams, ``Political polarization in the {A}merican
  public,'' \emph{Annu. Rev. Polit. Sci.}, vol.~11, pp. 563--588, 2008.

\bibitem{hetherington2009}
M.~J. Hetherington, ``Putting polarization in perspective,'' \emph{British
  Journal of Political Science}, vol.~39, no.~2, pp. 413--448, 2009.

\bibitem{hill2015}
S.~J. Hill and C.~Tausanovitch, ``A disconnect in representation? {C}omparison
  of trends in congressional and public polarization,'' \emph{The Journal of
  Politics}, vol.~77, no.~4, pp. 1058--1075, 2015.

\bibitem{mccarty2016}
N.~McCarty, K.~T. Poole, and H.~Rosenthal, \emph{Polarized America: The dance
  of ideology and unequal riches}.\hskip 1em plus 0.5em minus 0.4em\relax MIT
  Press, 2016.

\bibitem{abramowitz2005}
A.~Abramowitz and K.~Saunders, ``Why can’t we all just get along? {T}he
  reality of a polarized {A}merica,'' in \emph{The Forum}, vol.~3, no.~2.\hskip
  1em plus 0.5em minus 0.4em\relax Citeseer, 2005, pp. 1--22.

\bibitem{baldassarri2008}
D.~Baldassarri and A.~Gelman, ``Partisans without constraint: Political
  polarization and trends in {A}merican public opinion,'' \emph{American
  Journal of Sociology}, vol. 114, no.~2, pp. 408--446, 2008.

\bibitem{johnson1989formal}
P.~E. Johnson, ``Formal theories of politics: The scope of mathematical
  modelling in political science,'' in \emph{Formal Theories of
  Politics}.\hskip 1em plus 0.5em minus 0.4em\relax Elsevier, 1989, pp.
  397--404.

\bibitem{gersbach2017}
H.~Gersbach, P.~Muller, and O.~Tejada, ``A dynamic model of electoral
  competition with costly policy changes,'' \emph{CER-ETH--Center of Economic
  Research at ETH Zurich}, vol.~17, p. 270, 2017.

\bibitem{krasa2016political}
S.~Krasa, ``Political competition and the dynamics of parties and candidates,''
  2016, available at http://www.econ.uiuc.edu/~skrasa/party.pdf.

\bibitem{Yang2020}
V.~C. Yang, D.~M. Abrams, G.~Kernell, and A.~E. Motter, ``Why are {US} parties
  so polarized? {A} ``satisficing'' dynamical model,'' \emph{SIAM Review},
  vol.~62, no.~3, pp. 646--657, 2020.

\bibitem{Maximizing}
B.~Schwartz, A.~Ward, J.~Monterosso, S.~Lyubomirsky, K.~White, and D.~R.
  Lehman, ``Maximizing versus satisficing: happiness is a matter of choice.''
  \emph{Journal of personality and social psychology}, vol.~83, no.~5, p. 1178,
  2002.

\bibitem{jordan1998}
R.~Jordan, D.~Kinderlehrer, and F.~Otto, ``The variational formulation of the
  {F}okker--{P}lanck equation,'' \emph{SIAM journal on mathematical analysis},
  vol.~29, no.~1, pp. 1--17, 1998.

\bibitem{Villani2007}
C.~Villani, \emph{Optimal Transport: Old and New}, 1st~ed.\hskip 1em plus 0.5em
  minus 0.4em\relax Springer, Berlin, Heidelberg, 2007.

\bibitem{Santambrogio2015}
F.~Santambrogio, \emph{Optimal Transport for Applied Mathematicians}.\hskip 1em
  plus 0.5em minus 0.4em\relax Cham: Birkh{\"{a}}user, 2015.

\bibitem{Ambrosio2008}
L.~Ambrosio, N.~Gigli, and G.~Savar{\'{e}}, \emph{Gradient Flows},
  2nd~ed.\hskip 1em plus 0.5em minus 0.4em\relax Basel: Birkh{\"{a}}user, 2008.

\bibitem{santambrogio2017}
F.~Santambrogio, ``$\{$Euclidean, metric, and {W}asserstein$\}$ gradient flows:
  an overview,'' \emph{Bulletin of Mathematical Sciences}, vol.~7, no.~1, pp.
  87--154, 2017.

\bibitem{otto1996}
F.~Otto, \emph{Double degenerate diffusion equations as steepest
  descent}.\hskip 1em plus 0.5em minus 0.4em\relax Citeseer, 1996.

\bibitem{otto2001}
------, ``The geometry of dissipative evolution equations: the porous medium
  equation,'' \emph{Communications in Partial Differential Equations}, vol.~26,
  no. 1--2, pp. 101--174, 2001.

\bibitem{mei2019}
S.~Mei, T.~Misiakiewicz, and A.~Montanari, ``Mean-field theory of two-layers
  neural networks: dimension-free bounds and kernel limit,'' in
  \emph{Conference on Learning Theory}.\hskip 1em plus 0.5em minus 0.4em\relax
  PMLR, 2019, pp. 2388--2464.

\bibitem{bunne2021}
C.~Bunne, L.~Meng-Papaxanthos, A.~Krause, and M.~Cuturi, ``{JKO}net: Proximal
  optimal transport modeling of population dynamics,'' \emph{arXiv preprint
  arXiv:2106.06345}, 2021.

\bibitem{chewi2020}
S.~Chewi, T.~Maunu, P.~Rigollet, and A.~J. Stromme, ``Gradient descent
  algorithms for {B}ures-{W}asserstein barycenters,'' in \emph{Conference on
  Learning Theory}.\hskip 1em plus 0.5em minus 0.4em\relax PMLR, 2020, pp.
  1276--1304.

\bibitem{chizat2018}
L.~Chizat and F.~Bach, ``On the global convergence of gradient descent for
  over-parameterized models using optimal transport,'' \emph{Advances in Neural
  Information Processing Systems}, vol.~31, 2018.

\bibitem{zhang2018}
R.~Zhang, C.~Chen, C.~Li, and L.~Carin, ``Policy optimization as {W}asserstein
  gradient flows,'' in \emph{International Conference on Machine
  Learning}.\hskip 1em plus 0.5em minus 0.4em\relax PMLR, 2018, pp. 5737--5746.

\bibitem{richemond2017}
P.~H. Richemond and B.~Maginnis, ``On {W}asserstein reinforcement learning and
  the {F}okker-{P}lanck equation,'' \emph{arXiv preprint arXiv:1712.07185},
  2017.

\bibitem{arbel2019}
M.~Arbel, A.~Korba, A.~Salim, and A.~Gretton, ``Maximum mean discrepancy
  gradient flow,'' \emph{Advances in Neural Information Processing Systems},
  vol.~32, 2019.

\bibitem{salim2020}
A.~Salim, A.~Korba, and G.~Luise, ``The {W}asserstein proximal gradient
  algorithm,'' \emph{Advances in Neural Information Processing Systems},
  vol.~33, pp. 12\,356--12\,366, 2020.

\bibitem{lanzetti2022firstorder}
N.~Lanzetti, S.~Bolognani, and F.~Dörfler, ``First-order conditions for
  optimization in the {W}asserstein space,'' \emph{Working paper}, 2022.

\bibitem{dataa}
R.~Carroll, J.~Lewis, J.~Lo, N.~McCarty, K.~Poole, and H.~Rosenthal,
  ````\uppercase{C}ommon space'' {DW-NOMINATE} scores with bootstrapped
  standard errors (joint house and senate scaling),'' 2015, available at
  \url{http://voteview.com/dwnomin_joint_house_and_senate.htm}.

\bibitem{rudin1987}
W.~Rudin, \emph{Real and Complex Analysis}, 3rd~ed.\hskip 1em plus 0.5em minus
  0.4em\relax USA: McGraw-Hill, Inc., 1987.

\bibitem{bonnet2019}
B.~Bonnet, ``A {P}ontryagin {M}aximum {P}rinciple in {W}asserstein spaces for
  constrained optimal control problems,'' \emph{ESAIM: Control, Optimisation
  and Calculus of Variations}, vol.~25, p.~52, 2019.

\bibitem{bacsar1998dynamic}
T.~Ba{\c{s}}ar and G.~J. Olsder, \emph{Dynamic noncooperative game
  theory}.\hskip 1em plus 0.5em minus 0.4em\relax SIAM, 1998.

\bibitem{aolaritei2022}
L.~Aolaritei, N.~Lanzetti, H.~Chen, and F.~D{\"o}rfler, ``Uncertainty
  propagation via optimal transport ambiguity sets,'' \emph{arXiv preprint
  arXiv:2205.00343}, 2022.

\bibitem{Khalil2002}
H.~K. Khalil, \emph{Nonlinear Systems}, 3rd~ed.\hskip 1em plus 0.5em minus
  0.4em\relax Prentice Hall, 2002.

\end{thebibliography}

\end{document}